# Visualization System for Burrowing Organisms in Granular Substrates

Amos G. Winter, V
Anette E. Hosoi

Hatsopolous Microfluids Laboratory
Department of Mechanical Engineering
Massachusetts Institute of Technology

## Abstract

This fluid dynamics video demonstrates the capabilities of an experimental setup to visualize organisms burrowing in granular substrates. The setup consists of a tank filled with 1mm soda-lime glass beads, backlit by halogen lights. The walls can be moved such that the space bound by them can expand or contract to fit various organisms. A recirculation system oxygenates and cools the salt water that flows through the substrate. The video shows the burrowing behavior of a razor clam (*Ensis directus*) and a quahog (*Mercenaria mercenaria*). Both animals inflate their foot with blood, making the foot a terminal anchor against which they pull their shell downwards. Deformation of the granular substrate can be seen around the animals as they advance downwards.

The video can be downloaded from:

Small format
http://ecommons.library.cornell.edu/bitstream/1813/11477/3/Amos%20Winter_MIT_burrowing_small.mpg

Large format
http://ecommons.library.cornell.edu/bitstream/1813/11477/2/Amos%20Winter_MIT_burrowing_large.mpg